\documentclass[aps,prb,twocolumn,showpacs,showkeys,floatfix]{revtex4}

\usepackage{graphicx,color}
\usepackage{multirow,slashbox}

\newcommand{\jwj}[1]{\textcolor{black}{#1}}

\graphicspath{{figs/}}
\begin{document}

\title{Parametrization of Stillinger-Weber Potential Based on Valence Force Field Model: Application to Single-Layer MoS$_2$ and Black Phosphorus}

\author{Jin-Wu Jiang}
    \altaffiliation{Corresponding author: jiangjinwu@shu.edu.cn; jwjiang5918@hotmail.com}
    \affiliation{Shanghai Institute of Applied Mathematics and Mechanics, Shanghai Key Laboratory of Mechanics in Energy Engineering, Shanghai University, Shanghai 200072, People's Republic of China}

\date{\today}
\begin{abstract}

We propose to parametrize the Stillinger-Weber potential for covalent materials starting from the valence force field model. All geometrical parameters in the Stillinger-Weber potential are determined analytically according to the equilibrium condition for each individual potential term, while the energy parameters are derived from the valence force field model. This parametrization approach transfers the accuracy of the valence force field model to the Stillinger-Weber potential. Furthermore, the resulting Stilliinger-Weber potential supports for stable molecular dynamics simulations, as each potential term is at energy minimum state separately at the equilibrium configuration. We employ this procedure to parametrize Stillinger-Weber potentials for the single-layer MoS$_2$ and black phosphorous. The obtained Stillinger-Weber potentials predict accurate phonon spectrum and mechanical behaviors. We also provide input scripts of these Stillinger-Weber potentials used by publicly available simulation packages including GULP and LAMMPS.

\end{abstract}
\maketitle
\pagebreak

\section{Introduction}

\begin{figure}[tb]
  \begin{center}
    \scalebox{0.9}[0.9]{\includegraphics[width=8cm]{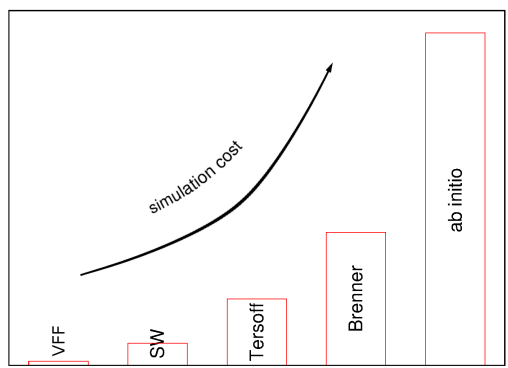}}
  \end{center}
  \caption{A schematic diagram comparing the simulation cost of different atomic interactions; i.e., VFF model, SW potential, Tersoff potential, Brenner potential, and {\it ab initio} approach.}
  \label{fig_compare_potential}
\end{figure}

The atomic interaction is a fundamental ingredient for numerical investigation of nearly all physical or mechanical processes. For instance, in molecular dynamics (MD) simulations, the atomic interaction provides the retracting force for each atom in the Newton's equation. There have been huge number of available potential models for the atomic interaction within different materials. For the covalent material, some representative potential models are shown in Fig.~\ref{fig_compare_potential} in the order of their simulation cost; i.e., valence force field (VFF) model, Stillinger-Weber (SW) potential, Tersoff potential, Brenner potential, and {\it ab initio} approaches. These potentials (or approaches) are able to describe the bond stretching and angle bending motions, which are two dominant motion styles in covalent materials. The bond twisting motion can also be treated by these potentials, although the twisting energy is usually very small.

The VFF model is a linear model, and is suitable for analytic derivation of many elastic quantities, so this model requires only limited computation cost. As an advantage of the VFF model, its parameters can be determined of high accuracy by fitting directly to some observable elastic quantities. As a result, the VFF model was very popular for covalent materials, especially before 1980s, when the CPU speed was very low. Consequently, the VFF model for most covalent materials have been well developed. For instance, the VFF model for MoS$_2$ has been proposed in 1975,\cite{WakabayashiN} while the VFF model for black phosphorus (BP) was proposed in 1982,\cite{KanetaC1982ssc} and the VFF model for graphene was developed in 1990 by Aizawa et al.\cite{AizawaT} These VFF models are useful for the study of many elastic properties in these quasi-two-dimensional nano-materials in recent years, especially during the gold rush of graphene in the past decade.

While the VFF model is beneficial for the fastest numerical simulation, its strong limitation is the absence of nonlinear effect. Due to this limitation, the VFF model is not applicable to nonlinear phenomena, for which other potential models with nonlinear components are required. The {\it ab initio} approach is accurate and applicable to nonlinear phenomena, but it requires the most expensive simulation cost, \jwj{due to the solution of the full quantum electronic problem.} However, this approach desires the most expensive simulation resources. As a result, the {\it ab initio} approach usually cannot simulate more than around a few thousand atoms, which poses serious limitations for comparisons to experimental studies.

We are now aware that the VFF model is the cheapest in computation cost, but it only works for elastic properties. On the other hand, the {\it ab initio} approach can simulate nearly all physical processes with high accuracy, but it requires the most expensive computation cost. Hence, the bridging between these two extreme cases is of practical significance, since lots of studies prefer efficient simulation with reasonable accuracy for the nonlinear treatment. There have been several potential forms to fill this bridging domain; including SW potential,\cite{StillingerF,JiangJW2013sw,JiangJW2014bpsw} Tersoff potential,\cite{TersoffJ1,TersoffJ3,TersoffJ2,TersoffJ4,TersoffJ5,JiangJW2011bngra,JiangJW2011bntube} and Brenner potential.\cite{brennerJPCM2002,LiangT,LindsayL2010prb} All of these potential forms comprise reasonable accurate nonlinear effects, and are particularly suitable for MD simulations.

Among these potentials, the SW potential is one of the simplest potential forms with nonlinear effects included.\cite{StillingerF} An advanced feature for the SW potential is that it includes the nonlinear effect, and keeps the numerical simulation at a very fast level. As a result, the SW potential has been widely used in the numerical simulation community. The SW potential was originally proposed by Stillinger and Weber to describe the interaction in solid and liquid forms of silicon, and it has been used in other covalent materials like single-layer MoS$_2$ (SLMoS$_2$)\cite{JiangJW2013sw} and single-layer BP (SLBP).\cite{JiangJW2014bpsw}

For chemically different materials, the SW potential form keeps unchanged, but all parameters need to be determined properly. In all present works, the parametrization of SW potential (and also Brenner and Tersoff potentials) are done by fitting to some experimentally known quantities like the Young's modulus, phonon spectrum, cohesion energy, and etc. Actually, from the above discussion, we have learnt that most covalent materials already have an accurate VFF model, which can describe linear properties accurately. Such attractive essence should be helpful for the parametrization of atomic potentials like SW potential, Tersoff potential, and Brenner potential. However, to-date, the accuracy of the VFF model was not transferred to other atomic potentials during their parametrization process. The present work takes the SW potential as an example to demonstrate the relationship between the VFF model and the SW potential. In doing so, we illustrate that the SW potential parameters can be analytically parametrized based on the VFF model.

In this paper, we propose a parametrization procedure for the development of SW potentials based on the VFF model. All SW geometrical parameters are determined according to the equilibrium condition for each SW term, while the SW energy parameters are derived from the VFF model analytically. This parametrization procedure is employed to develop the SW potentials for SLMoS$_2$ and SLBP, which provide accurate phonon spectrum and mechanical behaviors.

The present paper is organized as follows. In Sec.II, we present details about the parametrization of SW potential based on the VFF model. The parametrization procedure is applied to develop the SW potential for SLMoS$_2$ in Sec.III. Sec.IV is devoted to the analytic parametrization of the SW potential for the SLBP. The paper ends with a brief summary in Sec.V.

\section{VFF model and SW potential}

\begin{figure}[tb]
  \begin{center}
    \scalebox{1.1}[1.1]{\includegraphics[width=8cm]{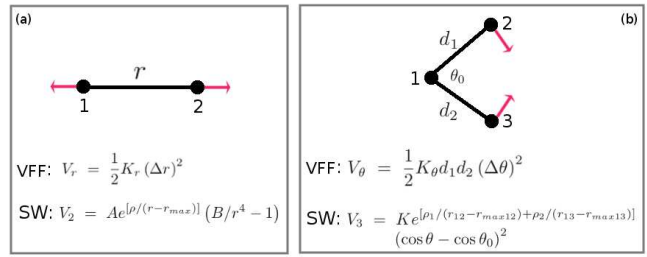}}
  \end{center}
  \caption{Two typical interactions in covalent materials. Each interaction term can be described using the VFF model or the SW potential. (a) Two-body bond stretching interaction. (b) Three-body angle bending interaction. Atom moving directions are depicted by red arrows.}
  \label{fig_vffm_sw}
\end{figure}

For most covalent bonding materials, the bond stretching and the angle bending are two typical motion styles as shown in Fig.~\ref{fig_vffm_sw}. The corresponding interactions can be described by the VFF model in the linear regime for small bond variation $\Delta r$ and angle variation $\Delta\theta$,
\begin{eqnarray}
V_{r} & = & \frac{1}{2}K_{r}\left(\Delta r\right)^{2},\\
\label{eq_vffm1}
V_{\theta} & = & \frac{1}{2}K_{\theta}d_{1}d_{2}\left(\Delta\theta\right)^{2},
\label{eq_vffm2}
\end{eqnarray}
where $K_{r}$ and $K_{\theta}$ are two VFF parameters. The $V_{r}$ term is the potential that captures a variation in the bond length $\Delta r$. The $V_{\theta}$ is for the potential corresponding to the variation of the angle $\Delta\theta$, where the anlge $\theta$ is formed by two bonds of length $d_{1}$ and $d_{2}$.

Besides VFF model, the SW potential is another useful potential for these two typical interactions in Fig.~\ref{fig_vffm_sw}. There are two-body and three-body interactions in the SW potential,
\begin{eqnarray}
V_{2} & = & Ae^{[\rho/\left(r-r_{max}\right)]}\left(B/r^{4}-1\right),
\label{eq_sw2}\\
V_{3} & = & Ke^{[\rho_{1}/\left(r_{12}-r_{max12}\right)+\rho_{2}/\left(r_{13}-r_{max13}\right)]}\left(\cos\theta-\cos\theta_{0}\right)^{2},\nonumber\\
\label{eq_sw3}
\end{eqnarray}
where $V_{2}$ corresponds to the bond stretching and $V_{3}$ associates with the angle bending. The cut-offs $r_{\rm max}$, $r_{\rm max12}$ and $r_{\rm max13}$ are geometrically determined by the material's structure. There are five unknown geometrical parameters, i.e., $\rho$ and $B$ in the two-body $V_2$ term and $\rho_1$, $\rho_2$, and $\theta_0$ in the three-body $V_3$ term, and two energy parameters $A$ and $K$.

Let's assume that the material's structure (bond length $d$ and angle $\theta_0$) has been identified via experiments or other accurate theoretical methods. Using these knowledge, we can determine geometrical parameters in the SW potential. First of all, it is reasonable to require that all bonds are at their equilibrium length and all angles are at their equilibrium angle value in the equilibrium configuration. That is, we have the equilibrium condition, $\frac{\partial V_{2}}{\partial r}|_{r=d}=0$ and $\frac{\partial V_{3}}{\partial\theta}|_{\theta=\theta_0}=0$, for each bond and each angle individually. From $\frac{\partial V_{2}}{\partial r}|_{r=d}=0$, we obtain the following constraint for parameters $\rho$ and $B$ in $V_{2}$, 
\begin{eqnarray}
\rho & = & \frac{-4B\left(d-r_{max}\right)^{2}}{\left(Bd-d^{5}\right)},
\label{eq_rho}
\end{eqnarray}
where $d$ is the equilibrium bond length from experiments. Hence, there is only one free geometrical parameter left in $V_2$. In other words, Eq.~(\ref{eq_rho}) ensures that the bond has an equilibrium length of $d$ and the $V_2$ interaction for this bond is at the energy minimum state at the equilibrium configuration.

The three-body $V_{3}$ term shown in Eq.~(\ref{eq_sw3}) ensures $\frac{\partial V_{3}}{\partial\theta}=0$ explicitly, so we have no constraint on geometrical parameters for the three-body term. In fact, there is no free geometrical parameter in $V_3$, because the angle $\theta_0$ is from the experiment while $\rho_1$ and $\rho_2$ have been determined by Eq.~(\ref{eq_rho}).

The energy parameters $A$ and $K$ in the SW potential can be derived from the VFF model, by equating the force constants from SW potential and the force constants in the VFF model. More specifically, we have $\frac{\partial^{2}V_{2}}{\partial r^{2}}|_{r=d}=K_{r}$ and $\frac{\partial^{2}V_{3}}{\partial\theta^{2}}|_{\theta=\theta_0}=K_{\theta}d_{1}d_{2}$ at the equilibrium structure, leading to,
\begin{eqnarray}
A & = & \frac{K_{r}}{\alpha e^{[\rho/\left(d-r_{max}\right)]}},
\label{eq_A}\\
K & = & \frac{K_{\theta}d_{1}d_{2}}{2\sin^{2}\theta_{0}e^{[\rho_{1}/\left(d_{1}-r_{\rm max12}\right)+\rho_{2}/\left(d_{2}-r_{\rm max13}\right)]}},
\label{eq_K}
\end{eqnarray}
where the coefficient $\alpha$ in Eq.~(\ref{eq_A}) is,
\begin{eqnarray}
\alpha & = & \left[\frac{\rho}{\left(d-r_{max}\right)^{2}}\right]^{2}\left(B/d^{4}-1\right)\nonumber\\
 & + & \left[\frac{2\rho}{\left(d-r_{max}\right)^{3}}\right]\left(B/d^{4}-1\right)\nonumber\\
 & + & \left[\frac{\rho}{\left(d-r_{max}\right)^{2}}\right]\left(\frac{8B}{d^{5}}\right)+\left(\frac{20B}{d^{6}}\right).
\end{eqnarray}
The bond length of the arms for the angle are $d_1$ and $d_2$, which are from experiments or other theoretical calculations. As a result, energy parameters in the SW potential are analytically related to the energy parameters in the VFF model.

We summarize the key steps in the above analytic parametrization of the SW potential. In the SW potential, bond stretching interaction is described by Eq.~(\ref{eq_sw2}), and angle bending interaction is described by Eq.~(\ref{eq_sw3}). The potential parameters are determined in three steps. First, interaction cut-offs ($r_{\rm max}$, $r_{\rm max12}$, and $r_{\rm max13}$) are determined geometrically by the equilibrium configuration of the material. The bond length ($d$, $d_1$, and $d_2$) and the angle ($\theta_0$) are also from the experiment or other theoretical calculations. Second, geometrical parameters $\rho$ in the two-body term and $\rho_1$ and $\rho_2$ in the three-body term are determined by Eq.~(\ref{eq_rho}), by assuming that each two-body SW term is at equilibrium separately. Third, energy parameters ($A$ and $K$) are determined by Eqs.~(\ref{eq_A}) and ~(\ref{eq_K}), based on the VFF model. In this way, we have analytically determined nearly all SW potential parameters uniquely, except the parameter $B$ for two-body SW potential in Eq.~(\ref{eq_sw2}). The above derivation shows that there is no constraint imposed on the parameter $B$ in the linear regime. The only condition for $B$ to satisfy is that $B<d^4$, so that $\rho>0$. We will explain in the next two sections that the parameter $B$ is related to the nonlinear mechanical process, and should be fixed according to a nonlinear quantity.

Before further processing, we note some advantages for the SW potential derived in this approach. First, such SW potential has fully inherited the accuracy of the VFF model, so it provides accurate description for linear properties which can be accurately described by the VFF model. Second, the equilibrium structure has been pre-built-in during the derivation as shown by Eq.~(\ref{eq_rho}), so this SW potential gives accurate relaxed configuration intrinsically. Third, each two-body and three-body term in the SW potential is fully relaxed separately at the equilibrium configuration; i.e., all bonds and angles are relaxed individually at the relaxed configuration. Hence, the SW potential will be extremely stable during MD simulations. Fourth, the SW potential includes nonlinear effects through the nonlinear forms of both two-body and three-body terms as shown in Eqs.~(\ref{eq_sw2}) and ~(\ref{eq_sw3}), so the SW potential is able to provide nonlinear properties, eg. via performing MD simulations.

\section{SW potential for MoS$_2$}

\begin{figure}[tb]
  \begin{center}
    \scalebox{0.5}[0.5]{\includegraphics[width=8cm]{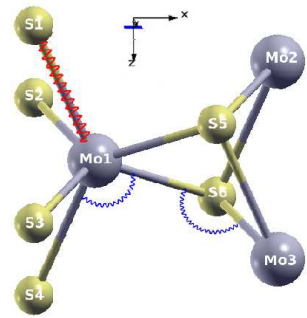}}
  \end{center}
  \caption{(Color online) Atomic configuration of SLMoS$_{2}$. There are two interaction types, i.e., the bond stretching term (red online) and the angle bending term (blue online). The x-axis is in the armchair direction, and the y-axis is in the zigzag direction.}
  \label{fig_cfg_mos2}
\end{figure}

\begin{table}
\caption{The VFF model parameters for SLMoS$_2$ from Ref~\onlinecite{WakabayashiN}.}
\label{tab_vffm_mos2}
\begin{tabular}{@{\extracolsep{\fill}}|c|c|c|}
\hline 
$K_{r}$ ($\frac{eV}{\AA^{2}}$) & $K_{\theta}$ ($\frac{eV}{\AA^{2}}$) & $K_{\psi}$ ($\frac{eV}{\AA^{2}}$)\tabularnewline
\hline 
\hline 
8.640 & 0.937 & 0.862\tabularnewline
\hline 
\end{tabular}
\end{table}

\begin{table}
\caption{Two-body (bond stretching) SW potential parameters for SLMoS$_2$ used by GULP. The expression is $V_{2}=Ae^{\left[\rho/\left(r-r_{max}\right)\right]}\left(B/r^{4}-1\right)$.}
\label{tab_sw2_gulp_mos2}
\begin{tabular}{@{\extracolsep{\fill}}|c|c|c|c|c|c|}
\hline 
 & $A$ (eV) & $\rho$ & $B$ (\AA$^4$) & $r_{\rm min} (\AA)$ & $r_{\rm max}$ (\AA)\tabularnewline
\hline 
\hline 
Mo-S & 6.918 & 1.252 & 17.771 & 0.0 & 3.16\tabularnewline
\hline 
\end{tabular}
\end{table}

\begin{table*}
\caption{Three-body (angle bending) SW potential parameters for SLMoS$_2$ used by GULP. The expression is $V_{3}=Ke^{\left[\rho_{1}/\left(r_{12}-r_{max12}\right)+\rho_{2}/\left(r_{13}-r_{max13}\right)\right]}\left(\cos\theta-\cos\theta_{0}\right)^{2}$. Mo-S-S indicates the bending energy for the angle with Mo as the apex.}
\label{tab_sw3_gulp_mos2}
\begin{tabular*}{\textwidth}{@{\extracolsep{\fill}}|c|c|c|c|c|c|c|c|c|c|c|}
\hline 
 & $K$ (eV) & $\theta_{0}$ (degree) & $\rho_{1}$ (\AA) & $\rho_{2}$ (\AA) & $r_{\rm min12}$ (\AA) & $r_{\rm max12}$ (\AA) & $r_{\rm min13}$ (\AA) & $r_{\rm max13}$ (\AA) & $r_{\rm min23}$ (\AA) & $r_{\rm max23}$ (\AA)\tabularnewline
\hline 
\hline 
Mo-S-S & 67.883 & 81.788 & 1.252 & 1.252 & 0.0 & 3.16 & 0.0 & 3.16 & 0.0 & 3.78\tabularnewline
\hline 
S-Mo-Mo & 62.449 & 81.788 & 1.252 & 1.252 & 0.0 & 3.16 & 0.0 & 3.16 & 0.0 & 4.27\tabularnewline
\hline 
\end{tabular*}
\end{table*}

\begin{figure}[tb]
  \begin{center}
    \scalebox{1.0}[1.0]{\includegraphics[width=8cm]{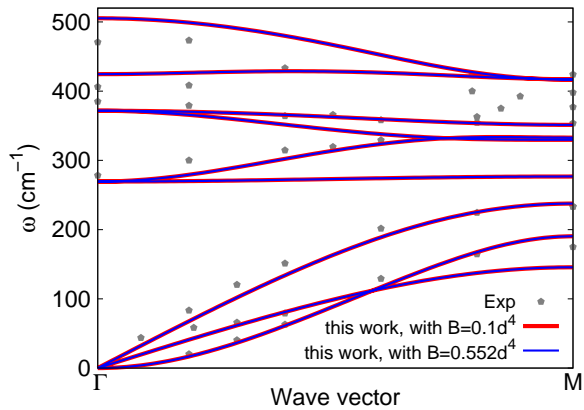}}
  \end{center}
  \caption{(Color online) Phonon spectrum for SLMoS$_{2}$ along the $\Gamma$M direction in the Brillouin zone. The results from the SW potential (lines) are compared with the experiment data (pentagons) from Ref~\onlinecite{WakabayashiN}. The parameter $B$ has no effect on the phonon spectrum.}
  \label{fig_phonon_mos2}
\end{figure}

\begin{figure}[tb]
  \begin{center}
    \scalebox{1}[1]{\includegraphics[width=8cm]{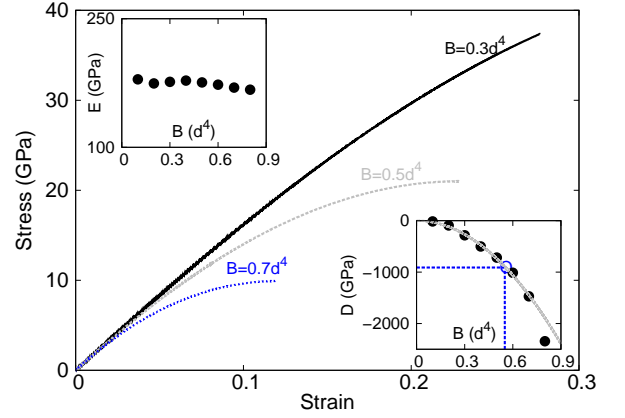}}
  \end{center}
  \caption{(Color online) The effect of parameter B on the stress-strain relation for SLMoS$_2$ of dimension $27.0\times 28.1$~{\AA} along the armchair direction at 1.0~K. The stress-strain curve is fitted to function $\sigma=E\epsilon+\frac{1}{2}D\epsilon^2$, with $E$ as the Young's modulus and $D$ as the TOEC. The left top inset shows that the parameter $B$ has no effect on the elastic property, Young's modulus; while the right bottom inset shows that the parameter $B$ dominates the nonlinear quantity, TOEC, which is fitted by function $D=-2953.8B^2$. The blue circle in the right bottom inset represents $D=-899.8$~{GPa} from the first-principles calculation,\cite{CooperRC2013prb1} which fixes parameter $B=0.552d^4$ for the SW potential.}
  \label{fig_toec_mos2}
\end{figure}

\begin{figure}[tb]
  \begin{center}
    \scalebox{1}[1]{\includegraphics[width=8cm]{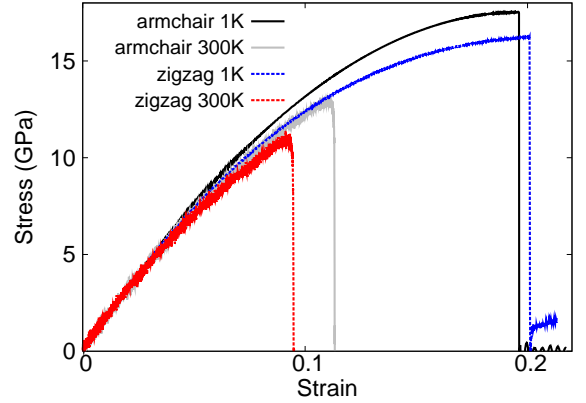}}
  \end{center}
  \caption{(Color online) Stress-strain for SLMoS$_2$ of dimension $27.0\times 28.1$~{\AA} along the armchair and zigzag directions. The Young's modulus is the same in the armchair and zigzag directions. The nonlinear mechanical properties are anisotropic in the armchair and zigzag directions.}
  \label{fig_stress_strain_mos2}
\end{figure}

\begin{table*}
\caption{SW potential parameters for SLMoS$_2$ used by LAMMPS.\cite{Lammps} The two-body potential expression is $V_{2}=\epsilon A\left(B_L\sigma^{p}r_{ij}^{-p}-\sigma^{q}r_{ij}^{-q}\right)e^{\left[\sigma\left(r_{ij}-a\sigma\right)^{-1}\right]}$. The three-body potential expression is $V_{3}=\epsilon\lambda e^{\left[\gamma\sigma\left(r_{ij}-a\sigma\right)^{-1}+\gamma\sigma\left(r_{jk}-a\sigma\right)^{-1}\right]}\left(\cos\theta_{jik}-\cos\theta_{0}\right)^{2}$. The quantity tol in the last column is a controlling parameter in LAMMPS.}
\label{tab_sw_lammps_mos2}
\begin{tabular*}{\textwidth}{@{\extracolsep{\fill}}|c|c|c|c|c|c|c|c|c|c|c|c|}
\hline 
 & $\epsilon$ (eV) & $\sigma$ (\AA) & $a$ & $\lambda$ & $\gamma$ & $\cos\theta_{0}$ & $A$ & $B_L$ & $p$ & $q$ & tol\tabularnewline
\hline 
\hline 
Mo-S-S & 1.000 & 1.252 & 2.523 & 67.883 & 1.000 & 0.143 & 6.918 & 7.223 & 4 & 0 & 0.0\tabularnewline
\hline 
S-Mo-Mo & 1.000 & 1.252 & 2.523 & 62.449 & 1.000 & 0.143 & 6.918 & 7.223 & 4 & 0 & 0.0\tabularnewline
\hline 
\end{tabular*}
\end{table*}

As an example, we apply the above parametrization procedure to develop the SW potential for SLMoS$_2$ in this section. We use the equilibrium structure for SLMoS$_2$ from the first-principles calculations as shown in Fig.~\ref{fig_cfg_mos2}. The bond length between neighboring Mo and S atoms is $d=2.382$~{\AA}, and the angles are $\theta=\angle SMoS=80.581^{\circ}$ and $\psi=\angle MoSMo=80.581^{\circ}$.

The VFF model for SLMoS$_2$ is from Ref~\onlinecite{WakabayashiN}, which is able to describe the phonon spectrum and the sound velocity accurately. We have listed the first three leading force constants for SLMoS$_2$ in Tab.~\ref{tab_vffm_mos2}, neglecting other weak interaction terms. The bond stretching term is $V_{r}=\frac{K_{r}}{2}\left(\Delta d\right)^{2}$ with $\Delta d$ as the length variation of Mo-S bond (eg. Mo$_1$-S$_1$). The angle bending term is $V_{\theta}=\frac{K_{\theta}}{2}d^{2}\left(\Delta\theta\right)^{2}$ for the angle Mo-S-S with Mo as the apex (eg. $\angle S_4Mo_1S_6$), and $V_{\psi}=\frac{K_{\psi}}{2}d^{2}\left(\Delta\psi\right)^{2}$ for angle S-Mo-Mo with S as the apex (eg. $\angle Mo_1S_6Mo_3$).

Using Eqs.~(\ref{eq_rho}), ~(\ref{eq_A}), and ~(\ref{eq_K}), we obtain the SW potential parameters for SLMoS$_2$ used by GULP\cite{gulp} as listed in Tabs.~\ref{tab_sw2_gulp_mos2} and ~\ref{tab_sw3_gulp_mos2}. We have found in Sec.II that the parameter B can not be determined by the linear VFF model, because B corresponds to the nonlinear mechanical behavior. In other words, parameter B has no effect on linear properties. For instance, we compute the phonon spectrum for the SLMoS$_2$ using two different sets of SW potential with $B=0.1d^4$ and $B=0.552d^4$. Although these two SW potential sets look completely different, Fig.~\ref{fig_phonon_mos2} shows that the phonon spectrum corresponding to different parameter $B$ are exactly the same.

To fix parameter $B$, a nonlinear quantity is needed. Fig.~\ref{fig_toec_mos2} clearly demonstrates that the parameter $B$ has strong effect on the nonlinear mechanical behavior of the stress-strain relation during the tension of a SLMoS$_2$ of dimension $27.0\times 28.1$~{\AA} at 1.0~K. The stress ($\sigma$) is fitted as a function of strain ($\epsilon$), $\sigma=E\epsilon+\frac{1}{2}D\epsilon^2$, with $E$ as the Young's modulus and $D$ as the third-order elastic constant (TOEC). The left top inset in Fig.~\ref{fig_toec_mos2} shows that the parameter $B$ has no effect on another elastic property, the Young's modulus. Fig.~\ref{fig_toec_mos2} right bottom inset shows the relationship between $D$ and parameter $B$. Using the first-principles result,\cite{CooperRC2013prb1} $D=-899.8$~{GPa}, we can fix the parameter $B=0.552d^4$.

The SW potential parameters for SLMoS$_2$ used by LAMMPS\cite{Lammps} are listed in Tab.~\ref{tab_sw_lammps_mos2}. The potential script for LAMMPS can be found in the supplemental material.\cite{SW2015sup} We use LAMMPS to perform MD simulations for the mechanical behavior of the SLMoS$_2$ under uniaxial tension at 1.0~K and 300.0~K. Fig.~\ref{fig_stress_strain_mos2} shows the stress-strain curve during the tension of a SLMoS$_2$ of dimension $27.0\times 28.1$~{\AA}. Periodic boundary conditions are applied in both armchair and zigzag directions. The structure is thermalized to the thermal steady state with the NPT (constant particle number, constant pressure, and constant temperature) ensemble for 100~ps by the Nos\'e-Hoover\cite{Nose,Hoover} approach. After thermalization, the MoS$_2$ is stretched in one direction at a strain rate of $10^8$~{s$^{-1}$}, while the stress in the lateral direction is allowed to be relaxed to be zero. We have used the inter-layer space in bulk MoS$_2$, 6.092~\AA, as the thickness of the SLMoS$_2$ in the computation of the strain energy density.

In Fig.~\ref{fig_stress_strain_mos2}, from the curve in the linear region, $\epsilon\in[0, 0.01]$, we get the Young's modulus of SLMoS$_2$ around 165.7~{GPa} and 167.0~{GPa} in the armchair and zigzag directions, respectively. \jwj{The shear modulus and Poisson's ratio can also be obtained in this linear regime.} It is obvious that the Young's modulus is isotropic for SLMoS$_2$ due to the three-fold rotational symmetry in this quasi hexagonal lattice structure.\cite{BornM} Recent experiments have measured the effective Young's modulus to be $E=120\pm 30$~{Nm$^{-1}$},\cite{CooperRC2013prb1,CooperRC2013prb2} or $E=180\pm 60$~{Nm$^{-1}$}.\cite{BertolazziS} These values correspond to an in-plane Young's modulus of $198.6\pm 49.7$~{GPa} or $297.9\pm 99.3$~{GPa}, considering the thickness of 6.092~{\AA}. Our theoretical values are quite close to the first experiment. The TOEC in the zigzag direction is larger than that in the armchair direction, which agrees with the first-principles calculations.\cite{CooperRC2013prb1} The SLMoS$_2$ yields at smaller strain at 300~K than 1.0~K for both armchair and zigzag directions.

In 2013, the author has parametrized with collaborators a SW potential set (SW2013-MoS$_2$) for the SLMoS$_2$ by fitting parameters to the experimental phonon spectrum.\cite{JiangJW2013sw} The present SW potential (SW2015-MoS$_2$) has fewer interaction components than the SW2013-MoS$_2$ potential. However, the phonon spectrum from SW2015-MoS$_2$ potential can be as accurate as the SW2013-MoS$_2$ potential, because the present parametrization procedure transfers the accuracy of the VFF model to the SW potential. Furthermore, each interaction component in the present SW2015-MoS$_2$ potential is at equilibrium invidually, which is more strict than the SW2013-MoS$_2$ potential, in which the equilibrium condition is satisfied overall among all interaction components. As a result, the SW2015-MoS$_2$ potential is more stable for MD simulations.

\section{SW potential for SLBP}

\begin{figure}[tb]
  \begin{center}
    \scalebox{0.7}[0.7]{\includegraphics[width=8cm]{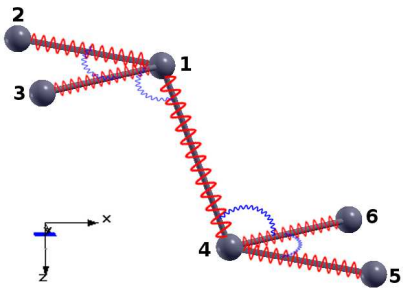}}
  \end{center}
  \caption{(Color online) Configuration of SLBP. Atoms are divided into the top group (atoms 1, 2, and 3) and the bottom group (atoms 4, 5, and 6). There are two interaction terms, the bond stretching term (red online) and the angle bending term (blue online). The x-axis is along the armchair direction, and the y-axis is along the zigzag direction.}
  \label{fig_cfg_bp}
\end{figure}

\begin{table}
\caption{The VFF model parameters for SLBP from Ref.\onlinecite{KanetaC1982ssc}.}
\label{tab_vffm_slbp}
\begin{tabular}{@{\extracolsep{\fill}}|c|c|c|}
\hline 
$K_{r}$ ($\frac{eV}{\AA^{2}}$) & $K_{\theta}$ ($\frac{eV}{\AA^{2}}$) & $K_{\psi}$ ($\frac{eV}{\AA^{2}}$)\tabularnewline
\hline 
\hline 
7.578 & 0.818 & 0.710\tabularnewline
\hline 
\end{tabular}
\end{table}

\begin{table}
\caption{Two-body (bond stretching) SW potential parameters for SLBP used by GULP. The expression is $V_{2}=Ae^{\left[\rho/\left(r-r_{max}\right)\right]}\left(B/r^{4}-1\right)$.}
\label{tab_sw2_gulp_slbp}
\begin{tabular}{@{\extracolsep{\fill}}|c|c|c|c|c|c|}
\hline 
 & $A$ (eV) & $\rho$ (\AA) & $B$ (\AA$^4$) & $r_{\rm min}$ (\AA) & $r_{\rm max}$ (\AA)\tabularnewline
\hline 
\hline 
P-P & 3.626 & 0.809 & 14.287 & 0.0 & 2.79\tabularnewline
\hline 
\end{tabular}
\end{table}

\begin{table*}
\caption{Three-body (angle bending) SW potential parameters for SLBP used by GULP. The expression is $V_{3}=Ke^{\left[\rho_{1}/\left(r_{12}-r_{max12}\right)+\rho_{2}/\left(r_{13}-r_{max13}\right)\right]}\left(\cos\theta-\cos\theta_{0}\right)^{2}$. The first two lines are for intra-group angles. The last two lines are for inter-group angles.}
\label{tab_sw3_gulp_slbp}
\begin{tabular*}{\textwidth}{@{\extracolsep{\fill}}|c|c|c|c|c|c|c|c|c|c|c|}
\hline 
 & $K$ (eV) & $\theta_{0}$ (degree) & $\rho_{1}$ (\AA) & $\rho_{2}$ (\AA) & $r_{\rm min12}$ (\AA) & $r_{\rm max12}$ (\AA) & $r_{\rm min13}$ (\AA) & $r_{\rm max13}$ (\AA) & $r_{\rm min23}$ (\AA) & $r_{\rm max23}$ (\AA)\tabularnewline
\hline 
\hline 
Pt-Pt-Pt & 35.701 & 96.359 & 0.809 & 0.809 & 0.0 & 2.79 & 0.0 & 2.79 & 0.0 & 3.89\tabularnewline
\hline 
Pb-Pb-Pb & 35.701 & 96.359 & 0.809 & 0.809 & 0.0 & 2.79 & 0.0 & 2.79 & 0.0 & 3.89\tabularnewline
\hline 
Pt-Pt-Pb & 32.006 & 102.094 & 0.809 & 0.809 & 0.0 & 2.79 & 0.0 & 2.79 & 0.0 & 3.89\tabularnewline
\hline 
Pb-Pb-Pt & 32.006 & 102.094 & 0.809 & 0.809 & 0.0 & 2.79 & 0.0 & 2.79 & 0.0 & 3.89\tabularnewline
\hline 
\end{tabular*}
\end{table*}

\begin{figure}[tb]
  \begin{center}
    \scalebox{1}[1]{\includegraphics[width=8cm]{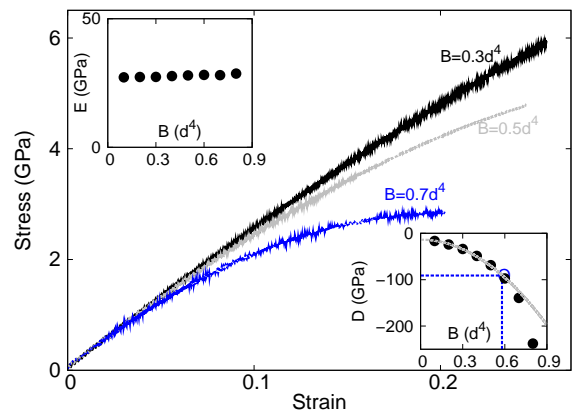}}
  \end{center}
  \caption{(Color online) The effect of parameter B on the stress-strain relation for SLBP along the armchair direction at 1.0~K. The stress-strain curve is fitted to function $\sigma=E\epsilon+\frac{1}{2}D\epsilon^2$, with $E$ as the Young's modulus and $D$ as the TOEC. Left top inset shows that parameter $B$ has no effect on the elastic quantity, Young's modulus. However, the right bottom inset shows that the parameter $B$ has strong effect on the nonlinear property, TOEC, which is fitted to function $D=-13.8-227.1B^2$. The blue circle in the right bottom inset represents $D=-91.3$~{GPa} from the first-principles calculation,\cite{WeiQ2014apl} which helps to fix parameter $B=0.584d^4$ for the SW potential.}
  \label{fig_toec_bp}
\end{figure}

\begin{figure}[tb]
  \begin{center}
    \scalebox{1}[1]{\includegraphics[width=8cm]{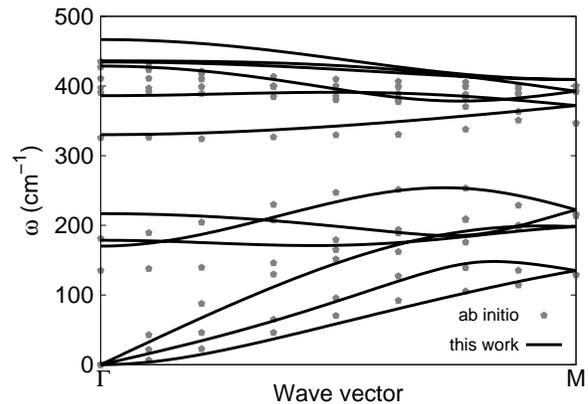}}
  \end{center}
  \caption{(Color online) Phonon spectrum for SLBP along $\Gamma$M from the SW potential is compared to the data from the {\it ab initio} calculation.\cite{ZhuZ2014prl}.}
  \label{fig_phonon_bp}
\end{figure}

\begin{figure}[tb]
  \begin{center}
    \scalebox{1}[1]{\includegraphics[width=8cm]{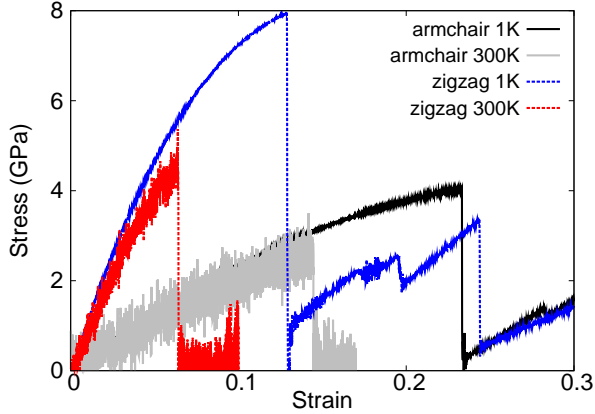}}
  \end{center}
  \caption{(Color online) Stress-strain for SLBP during tension process. Highly anisotropic mechanical behaviors are observed in the armchair and zigzag directions.}
  \label{fig_stress_strain_bp}
\end{figure}

\begin{table*}
\caption{SW potential parameters for SLBP used by LAMMPS. The two-body potential expression is $V_{2}=\epsilon A\left(B_L\sigma^{p}r_{ij}^{-p}-\sigma^{q}r_{ij}^{-q}\right)e^{\left[\sigma\left(r_{ij}-a\sigma\right)^{-1}\right]}$. The three-body potential expression is $V_{3}=\epsilon\lambda e^{\left[\gamma\sigma\left(r_{ij}-a\sigma\right)^{-1}+\gamma\sigma\left(r_{jk}-a\sigma\right)^{-1}\right]}\left(\cos\theta_{jik}-\cos\theta_{0}\right)^{2}$. The quantity tol in the last column is a controlling parameter in LAMMPS. Pt indicates atoms from the top group, while Pb represents atoms in the bottom group.}
\label{tab_sw_lammps_slbp}
\begin{tabular*}{\textwidth}{@{\extracolsep{\fill}}|c|c|c|c|c|c|c|c|c|c|c|c|}
\hline 
 & $\epsilon$ (eV) & $\sigma$ (\AA) & $a$ & $\lambda$ & $\gamma$ & $\cos\theta_{0}$ & $A$ & $B_L$ & $p$ & $q$ & tol\tabularnewline
\hline 
\hline 
Pt-Pt-Pt & 1.000 & 0.809 & 3.449 & 35.701 & 1.000 & -0.111 & 3.626 & 33.371 & 4 & 0 & 0.0\tabularnewline
\hline 
Pb-Pb-Pb & 1.000 & 0.809 & 3.449 & 35.701 & 1.000 & -0.111 & 3.626 & 33.371 & 4 & 0 & 0.0\tabularnewline
\hline 
Pt-Pt-Pb & 1.000 & 0.809 & 3.449 & 32.006 & 1.000 & -0.210 & 0.000 & 33.371 & 4 & 0 & 0.0\tabularnewline
\hline 
Pb-Pb-Pt & 1.000 & 0.809 & 3.449 & 32.006 & 1.000 & -0.210 & 0.000 & 33.371 & 4 & 0 & 0.0\tabularnewline
\hline 
\end{tabular*}
\end{table*}

As another example, we apply the parametrization procedure to develop the SW potential for SLBP in this section. The structure for SLBP shown in Fig.~\ref{fig_cfg_bp} has been identified by experiment.\cite{TakaoY1981physica} P atoms are divided into the top group (including atoms 1, 2, and 3) and the bottom group (including atoms 4, 5, and 6). There are two bond lengths, i.e., the intra-group bond (eg. bond 1-2) $d_{1}=2.224$~{\AA} and the inter-group bond (eg. bond 1-4) $d_{2}=2.244$~{\AA}. These two bond lengths are very close to each other, so it can be assumed that both bonds have the same length of\cite{KanetaC1982ssc} $d=2.224$~{\AA}. The intra-group angle (eg. $\angle 213$) is $\theta=96.359^{\circ}$ and the inter-group angle (eg. $\angle 314$) is $\psi=102.09^{\circ}$.

Tab.~\ref{tab_vffm_slbp} lists the VFF model parameters for SLBP from Ref.\onlinecite{KanetaC1982ssc}. The bond stretching potential between two neighboring P atoms is $V_{r}=\frac{K_{r}}{2}\left(\Delta d\right)^{2}$. We note that the intra-group bond and the inter-group bond essentially have the same stretching parameter.\cite{KanetaC1982ssc} As a result, there is only one VFF model parameter for bond stretching potential. The angle bending potential is $V_{\theta}=\frac{K_{\theta}}{2}d^{2}\left(\Delta\theta\right)^{2}$ for the intra-group angle, and $V_{\psi}=\frac{K_{\psi}}{2}d^{2}\left(\Delta\psi\right)^{2}$ for the inter-group angle. These three terms make dominant contribution to the interaction for the SLBP, while other weak interaction terms have been omitted in the present work. As a compensate, these parameters in Tab.~\ref{tab_vffm_slbp} are different from the original value by an overall factor of 0.76.

Using Eqs.~(\ref{eq_rho}), ~(\ref{eq_A}), and ~(\ref{eq_K}), we obtain the SW potential parameters for SLBP used by GULP\cite{gulp} as shown in Tabs.~\ref{tab_sw2_gulp_slbp} and ~\ref{tab_sw3_gulp_slbp}. The determination of $B$ is illustrated in Fig.~\ref{fig_toec_bp}. The parameter $B$ has no effect on the elastic property, the Young's modulus, as shown by the left top inset in Fig.~\ref{fig_toec_bp}. However, the parameter $B$ has strong effect on the nonlinear quantity, TOEC, which can be fitted to the function $D=-13.8-227.1B^2$. Using this relationship between the TOEC and parameter $B$, we obtain the parameter $B=0.584d^4$ corresponding to $D=-91.3$~{GPa} from the first-principles calculations.\cite{WeiQ2014apl} We note that $D=-13.8 \not=0$ even for $B=0$, as shown in the right bottom inset of Fig.~\ref{fig_toec_bp}. For $B=0$, the only nonzero SW potential term is $V_3=K(\cos\theta-\cos\theta_0)^2$, so the nonzero residue, $D=-13.8$~{GPa}, originates from the nonlinear effect purely contributed by the angle bending interaction. This is different from SLMoS$_2$ results shown in the right bottom inset in Fig.~\ref{fig_toec_mos2}, where $D=0$ at $B=0$. This difference can be attributed to the different space groups for SLBP (C$_{2h}$) and SLMoS$_2$ (D$_{3h}$). As a restriction of the three-fold symmetry in the SLMoS$_2$, the overall nonlinear effect from the angle bending vanishes.

The phonon spectrum for the SLBP from the SW potential is shown in Fig.~\ref{fig_phonon_bp}. The results from SW potential agrees quite well with the first-principles calculations.\cite{ZhuZ2014prl}

SW potential parameters for SLBP used by LAMMPS\cite{Lammps} are listed in Tab.~\ref{tab_sw_lammps_slbp}. The potential script for LAMMPS can be found in the supplemental material.\cite{SW2015sup} We use LAMMPS to perform MD simulations for the tensile behavior for the SLBP of dimension $26.3\times 29.8$~{\AA} at 1.0~K and 300.0~K. Fig.~\ref{fig_stress_strain_bp} shows the stress-strain curves during the tensile deformation of the SLBP along the armchair direction and the zigzag direction. Periodic boundary conditions are applied in both armchair and zigzag directions. The structure is thermalized to the thermal steady state with the NPT (constant particle number, constant pressure, and constant temperature) ensemble for 100~ps by the Nos\'e-Hoover\cite{Nose,Hoover} approach. After thermalization, the SLBP is stretched in one direction at a strain rate of $10^8$~{s$^{-1}$}, and the stress in the lateral direction is allowed to be fully relaxed. We have used the inter-layer space of 5.24~$\AA$ as the thickness of the SLBP in the computation of the strain energy density.

In Fig.~\ref{fig_stress_strain_bp}, from the stress-strain curve in the strain range [0, 0.01], we obtain the Young's modulus 33.5~{GPa} and 105.5~{GPa} in the armchair and zigzag directions, respectively. These values are close to the previously reported {\it ab initio} results, eg. 28.9~{Nm$^{-1}$} in the armchair direction and 101.6~{Nm$^{-1}$} in the zigzag direction from Ref.\onlinecite{QiaoJ2014nc}. The SLBP yields at smaller strain at 300~K than 1.0~K for both armchair and zigzag directions.

In a recent work, the author has parametrized with collaborators a SW potential set (SW2013-BP) for the SLBP by fitting parameters to the phonon spectrum from {\it ab initio} calculations.\cite{JiangJW2014bpsw} The present SW potential (SW2015-BP) has fewer interaction components than the SW2013-BP potential. However, the phonon spectrum from SW2015-BP potential can be as accurate as the SW2013-BP potential, because the present parametrization procedure transfers the accuracy of the VFF model to the SW potential. Furthermore, each interaction component in the present SW2015-BP potential is at equilibrium invidually, which is more strict than the SW2013-BP potential, in which the equilibrium condition is satisfied overall among all interaction components. As a result, the SW2015-BP potential is more stable for MD simulations.

As a final note, this work proposes a method to develop the SW potential based on the VFF model, and applies this parametrization approach to SLMoS$_2$ and SLBP. The parametrization procedure, represented in Sec.II, is actually applicable to the development of other atomic potentials for a wide range of covalent materials. It is quite obvious that the SW potential for other covalent materials can also be developed analogously.

\textbf{An important technical note. For the simulation of SLMoS$_2$ by LAMMPS, one needs to recompile the LAMMPS package with our modified source file, $pair\_sw.cpp$, in the supplemental material.\cite{SW2015sup} This helps to exclude angle bending for angles like $\angle S_1 Mo_1 S_4$ in Fig.~\ref{fig_cfg_mos2}, which is not considered in the present work. However, for the simulation of SLBP using LAMMPS, one must use the original LAMMPS package; i.e., use the original source file, $pair\_sw.cpp$.}

\section{conclusion}

In conclusion, we have proposed an approach to determine the SW potential parameters based on the valence force field model. The SW potential developed following this approach inherits the accuracy of the VFF model in the description of linear physical properties. Furthermore, the accurate equilibrium structure information is pre-built-in, and this potential is very suitable for stable MD simulations. Finally, the SW potential can be easily used in many available MD simulation packages such as GULP and LAMMPS. As two examples, we apply this parametrization technique to develop the SW potential for SLMoS$_2$ and SLBP, which are found to provide accurate phonon spectrum and mechanical properties.


\textbf{Acknowledgements} The author thanks R. Timon and Harold S. Park for comments. The work is supported by the Recruitment Program of Global Youth Experts of China and the start-up funding from Shanghai University.



%
\end{document}